\documentclass[amsmath, amsthm, amssymb, floatfix, reprint, onecolumn, notitlepage, nofootinbib, longbibliography, 12pt]{revtex4-2}
\usepackage{setspace}
\usepackage{amsmath}
\usepackage{breqn}
\usepackage{graphicx}
\usepackage[nearskip,margin = 0pt,caption=false]{subfig}

\usepackage{verbatim}
\usepackage{amsfonts}
\usepackage{amssymb}
\usepackage{epstopdf}
\usepackage{lipsum}
\usepackage{siunitx}
\usepackage{xcolor}
\usepackage{array}
\usepackage{braket}
\usepackage{booktabs}
\usepackage[normalem]{ulem}
\usepackage[resetlabels,labeled]{multibib}
\DeclareGraphicsExtensions{.pdf,.eps,.png,.jpg,.mps}
\usepackage{etoolbox}
\usepackage{subfiles} 
\renewcommand{\baselinestretch}{2.0}

\DeclareUnicodeCharacter{0327}{ }

\begin{document}
\title{Unified laser stabilization and isolation on a silicon chip
}
\baselineskip=12pt
\author{Alexander D. White$^{1, \dagger, *}$, Geun Ho Ahn$^{1, \dagger, *}$, Richard Luhtaru$^{1}$, Joel Guo$^2$, Theodore J. Morin$^2$, Abhi Saxena$^3$, Lin Chang$^2$, Arka Majumdar$^3$, Kasper Van Gasse$^{1}$, John E. Bowers$^{2}$, Jelena Vu\v{c}kovi\'{c}$^{1}$\\
\vspace{+0.00 in}
{\small
$^1$E. L. Ginzton Laboratory, Stanford University, Stanford, CA 94305, USA.\\
\vspace{-0.1 in}
$^2$ ECE Department, University of California, Santa Barbara, CA, 93106, USA\\
\vspace{-0.1 in}
$^3$ ECE Department, University of Washington, Seattle, WA, 98195, USA\\
\vspace{-0.05 in}
$^{\dagger}$These authors contributed equally to this work.}\\
{\small $*$ adwhite@stanford.edu, gahn@stanford.edu}}
\baselineskip=24pt


\begin{abstract}
    \renewcommand{\baselinestretch}{2.0}
    \noindent \textbf{Rapid progress in photonics has led to an explosion of integrated devices that promise to deliver the same performance as table-top technology at the nanoscale; heralding the next generation of optical communications, sensing and metrology, and quantum technologies. 
    However, the challenge of co-integrating the multiple components of high-performance laser systems has left application of these nanoscale devices thwarted by bulky laser sources that are orders of magnitude larger than the devices themselves. 
    Here we show that the two main ingredients for high-performance lasers -- noise reduction and isolation -- currently requiring serial combination of incompatible technologies, can be sourced simultaneously from a single, passive, CMOS-compatible nanophotonic device.  
    To do this, we take advantage of both the long photon lifetime and the nonreciprocal Kerr nonlinearity of a high quality factor silicon nitride ring resonator to self-injection lock a semiconductor laser chip while also providing isolation. 
    Additionally, we identify a previously unappreciated power regime limitation of current on-chip laser architectures which our system overcomes. 
    Using our device, which we term a unified laser stabilizer, we demonstrate an on-chip integrated laser system with built-in isolation and noise reduction that operates with turnkey reliability.
    This approach departs from efforts to directly miniaturize and integrate traditional laser system components and serves to bridge the gap to fully integrated optical technologies.
    }
\end{abstract}

\maketitle
\renewcommand{\baselinestretch}{2.0}


\vspace{30pt}



Coherent optical sources serve as the backbone of optical communication, sensing and metrology, and quantum technologies \cite{kikuchi2015fundamentals, mcgrew2018atomic, daley2022practical}. Long coherence times and narrow linewidths allow for more precise heterodyne detection, interferometry, and probing of atomic transitions. Traditionally, the narrow linewidth optical sources required for these applications are built from table-top laser systems, but technological innovation in optical networking \cite{yang2022multi, rizzo2023massively}, data processing \cite{feldmann2021parallel,ashtiani2022chip, pai2023experimentally}, light detection and ranging (LiDAR) \cite{trocha2018ultrafast,rogers2021universal, li2023frequency}, and chip scale quantum computers \cite{wang2020integrated, madsen2022quantum, zheng2023multichip} demand the same performance from compact and fully integrated systems. In this vein, significant efforts have been made to miniaturize and integrate the components of low-noise lasers.
To do this, three key components must be combined: a semiconductor gain chip or laser, an external cavity for linewidth stabilization, and an optical isolator to prevent unwanted reflections from destabilizing the laser. 



Integrated III-V semiconductor lasers form the backbone of the internet, and their power, linewidth and stability have been continuously improving over the last few decades. However, to further improve linewidth and to increase tunability and stability, it is necessary to couple lasers to additional photonic components. While III-V materials like InP and GaAs serve as excellent gain media, their processing complexity leads to large waveguide losses and incompatibility with complementary metal–oxide–semiconductor (CMOS) processing, rendering monolithic integration of photonic devices in III-V undesirable \cite{zhou2023prospects}. Additionally, while some types of semiconductor lasers are less sensitive to back-reflection \cite{liu2017reflection,dong2021dynamic,matsui2021low}, they are still in the regime where optical isolators are critical for preventing destabilization \cite{tkach1986regimes}.

Recently, extremely high quality factor (Q $> \num{2e8}$) silicon nitride ring cavities have been used in conjunction with chip-scale semiconductor lasers to achieve Hz level linewidths on an integrated platform \cite{jin2021hertz, guo2022chip}, rivaling even the highest performance table-top systems. These devices work by taking advantage of the naturally occurring back-scattering in high Q resonators to provide narrow linewidth feedback that self-injection locks a semiconductor laser. As back-scattering is nearly ubiquitous in high Q resonators, this approach is reliable even though it is not deterministic. However, as we will show in this paper, when power levels in the device are high enough to access nonlinear processes in the resonator material, this technique starts to break down. This effect imposes stringent limits on the maximum operating power and quality factor of current architectures. Additionally, these devices still require external isolators. 

Integrated isolators have also made significant strides recently, with demonstrations of on-chip isolators approaching the performance of stand-alone optical elements \cite{herrmann2022mirror, yu2023integrated}. However, there are significant challenges in integrating these with low-noise systems. For instance, resonant devices \cite{tian2021magnetic, herrmann2022mirror} would require the tuning of multiple high Q resonators to degenerate frequencies. Meanwhile, nonresonant devices can be quite large ($>$ 1~cm) and, without a resonant filter, only provide a frequency shift of the back-reflected power reaching the laser cavity \cite{yu2023integrated}. While this has been shown to prevent laser destabilization, it is not clear that it would allow for the preservation of ultra-narrow linewidth. 

Despite this great progress in the integration of individual laser components, there is not a clear path to co-integrate these components, which are built using disparate and often incompatible technological platforms.
As a result, fully integrated laser systems remain elusive. This necessitates driving photonic integrated circuits with external sources, often orders of magnitude larger than the circuits themselves \cite{yang2022multi, rizzo2023massively, feldmann2021parallel,ashtiani2022chip, pai2023experimentally,trocha2018ultrafast,rogers2021universal, li2023frequency,wang2020integrated, madsen2022quantum, zheng2023multichip}.

Here we propose and demonstrate a single CMOS-compatible device, which we term a unified laser stabilizer (ULS), that passively and simultaneously feedback-stabilizes and isolates semiconductor lasers. 
To do this, we use the high quality factor of a ring resonator as a resource for both long photon lifetimes and optical nonlinearity. Long photon lifetimes allow us to generate linewidth reduction through self-injection locking, and optical nonlinearity allows us to generate optical isolation through the nonreciprocal Kerr effect.
Operating the ring resonator as a circulator, we can deterministically provide strong feedback (and thus large linewidth reductions) independent of input power, enabling concurrent linewidth narrowing and isolation. 
As the isolation and feedback come from the same ring, the laser is always on resonance, relaxing system complexity and enabling turnkey operation.

\begin{figure}[h!]
\centering\includegraphics[width=1\linewidth]{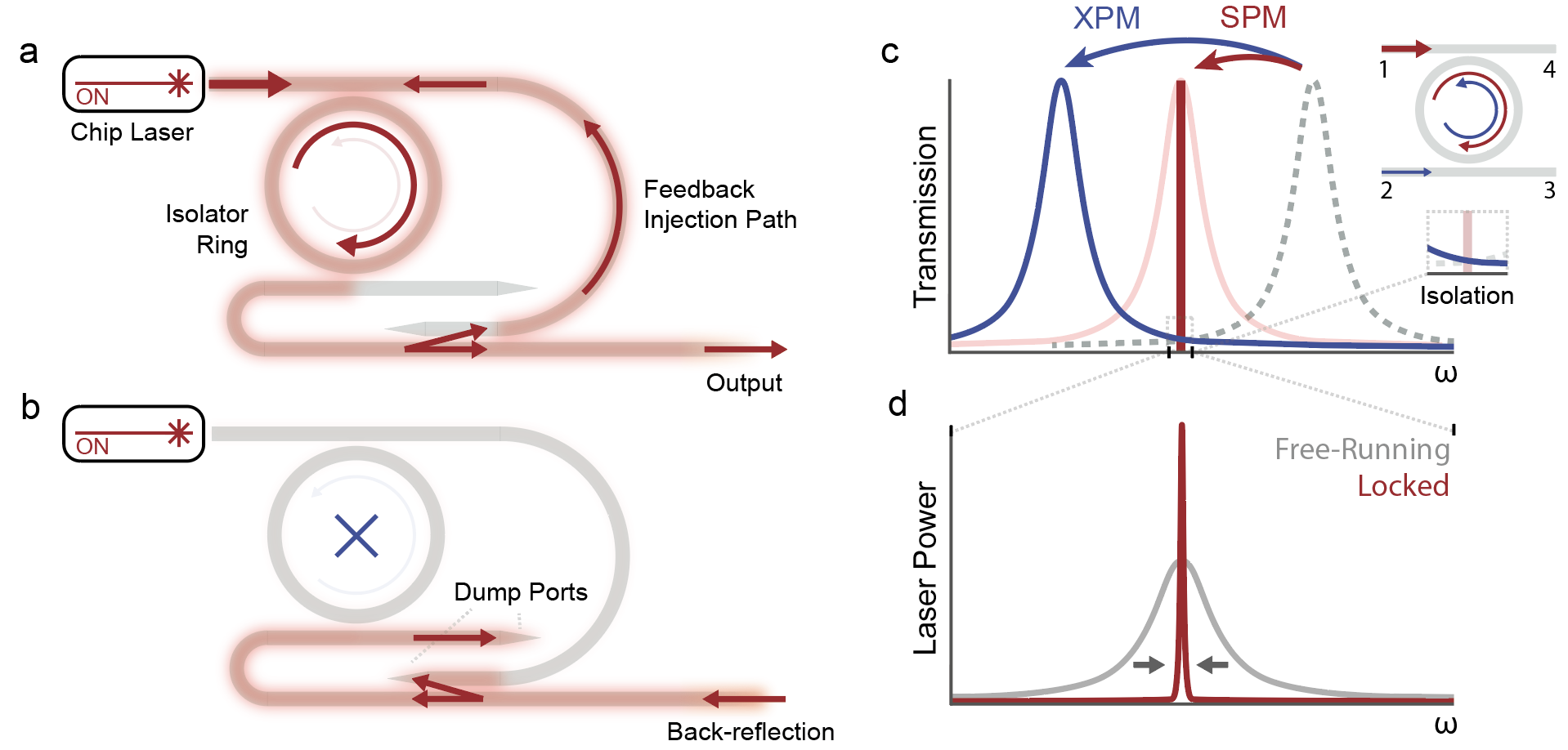}
\caption{{\bf Theory of Operation.} {\bf a}. Device schematic. Chip laser pumps isolator ring, whose output is tapped off to provide feedback to the laser. As the ring is being pumped in the clockwise direction, the power traveling from the feedback injection path to the laser is not resonant with the counterclockwise mode of the ring and travels back fully into the laser, stabilizing it. 
{\bf b}. Device under the influence of a back-reflection. Back-reflected power (in a frequency band near the pump) is not resonant with the counterclockwise mode of the ring and gets dumped, unable to reach the laser. 
{\bf c}. Transmission spectrum of isolator ring in the clockwise (red) and counterclockwise (blue) modes. Dashed line shows the degenerate cold cavity spectrum of the ring. This splitting is due to the 2-times difference in strength between self-phase and cross-phase modulation in the ring. 
{\bf d}. Effect of the feedback on the laser linewidth. }
\label{fig:Fig1}
\end{figure}

\section{Theory of Operation}



High quality factor ring resonators can be used to directly isolate the output of a continuous wave (CW) laser \cite{del2017symmetry, del2018microresonator, white2023integrated}. This works by taking advantage of the nonreciprocity of the optical Kerr effect (third-order nonlinearity $\chi^{(3)}$). When one of the degenerate clockwise and counterclockwise modes of a ring is pumped, the differential action of self-phase (SPM) and cross-phase modulation (XPM) induce a differential shift in the clockwise and counterclockwise resonance frequencies. As XPM is twice as strong as SPM, the ring mode counter-propagating with respect to the pump is shifted twice as far in frequency as the pump mode, leading to the split transmission spectrum illustrated in Fig 1c. The pump can then transmit through the ring with near unity efficiency, while any power reflected back is no longer resonant with the ring, and is thus isolated.

The maximum isolation achieved with this scheme can be found by taking the Lorentzian transmission of the detuned resonance \cite{del2018microresonator, white2023integrated}: 
\begin{align}
    I &= \frac{1}{1 + (2Q\frac{\Delta\omega}{\omega_0})^2}, \\
    \Delta\omega &= \omega_0 \frac{n_2}{n} \frac{Q \lambda}{2\pi V_{\text{mode}}}  \eta P_{\text{in}},
\end{align}
where $\omega_0$ is the original resonance frequency of the ring, $n_2$ is the nonlinear refractive index, $n$ is the linear refractive index, $Q$ is the loaded quality factor of the ring, $V_{\text{mode}}$ is the mode volume of the ring, $P_{\text{in}}$ is the input power, and $\eta$ is the coupling efficiency of the pump to the ring. At the peak of the forward resonance, where isolation is maximized, the coupling efficiency $\eta = \frac{4\kappa_1 (\kappa_2 + \gamma)}{(\kappa_1 + \kappa_2 + \gamma)^2}$ depends on the coupling rates $\kappa_1, \kappa_2$ and the intrinsic loss rate $\gamma$.
As the isolation ratio scales as $Q^4$, it is highly desirable to operate with a large quality factor ($> 10^6$). This not only ensures the maximum possible isolation, but also reduces the power threshold required to induce isolation.

While it serves to provide isolation, the high quality factor ring can also be utilized as a resource for frequency stability.
This is achieved by coupling the semiconductor laser to the high Q resonator through feedback, effectively creating an external cavity laser \cite{jin2021hertz, guo2022chip}. As the fundamental coherence limit of a laser, given by the Schawlow-Townes linewidth \cite{schawlow1958infrared}
\begin{equation}
    \Delta\nu_{laser} \geq  \frac{\pi h \nu (\Delta \nu_{\text{cav}})^2}{P_{\text{out}}} = \frac{\pi h \nu^3}{Q^2 P_{\text{out}}},
\end{equation}
(where $h$ is the Planck constant, $\nu$ is the laser frequency, and $P_{out}$ is the laser power) is dependent on Q, the fundamental frequency noise of a laser can be reduced by the introduction of a high Q feedback by the noise reduction factor (NRF)
\begin{equation}
    \text{NRF} \propto \Gamma \frac{(Q_{\text{cav}})^2}{(Q_{\text{laser}})^2},
\end{equation}
where $\Gamma$ is the fraction of power fed back to the laser. As semiconductor chip lasers typically have a low quality factor ($Q_{\text{laser}} \sim 10^4$) to maximize their output power, we can use the high quality factor rings ($Q_{\text{cav}} > 10^6$) to reduce their frequency noise by orders of magnitude. In the remainder of the paper, we use Q to denote the loaded cavity Q factor (labeled as $Q_{cav}$ above). 


To achieve simultaneous isolation and feedback stabilization, we propose the topology illustrated in Fig 1a, which we term a unified laser stabilizer. A chip laser is coupled to an isolator ring, which transmits power in the clockwise mode to an output waveguide. This output waveguide is tapped off with a directional coupler, and a fraction of the power is fed back to the laser. To reach the laser, the feedback power has to pass by the ring. As it is now traveling in the opposite direction, it interacts only with the detuned counterclockwise mode, preventing power coupling to the ring and allowing full transmission back to the laser. 
Meanwhile, any back-reflected power is off-resonant with the ring and gets dumped, preventing it from reaching the laser diode (Fig 1b).
Here the ring acts effectively as a circulator. If the ring is critically coupled ($\kappa_1 = \kappa_2 \gg \gamma$), transmission is only allowed from ports $1 \rightarrow 2 \rightarrow 3 \rightarrow 4 \rightarrow 1$ (as labeled in Fig 1c). If the ring is not critically coupled, the allowed transmission is $1 \rightarrow 2 \leftrightarrow 3 \rightarrow 4 \leftrightarrow 1$. 
The circulator behavior allows a large fraction of the output power to be fed back to the laser diode, leading to a strong reduction in the laser linewidth (Fig 1d). 

To optimize the isolation and noise reduction of a unified laser stabilizer, there are several important design considerations; most notably ring loading, resonator mode volume, and feedback strength.
Of course, as isolation and noise reduction factor scale as a function of $Q^4$ and $Q^2$ respectively, improving $Q$ will greatly improve the all-around performance. Given a fixed intrinsic Q, there are direct tradeoffs.

\begin{figure}[b!]
\centering\includegraphics[width=0.95\linewidth]{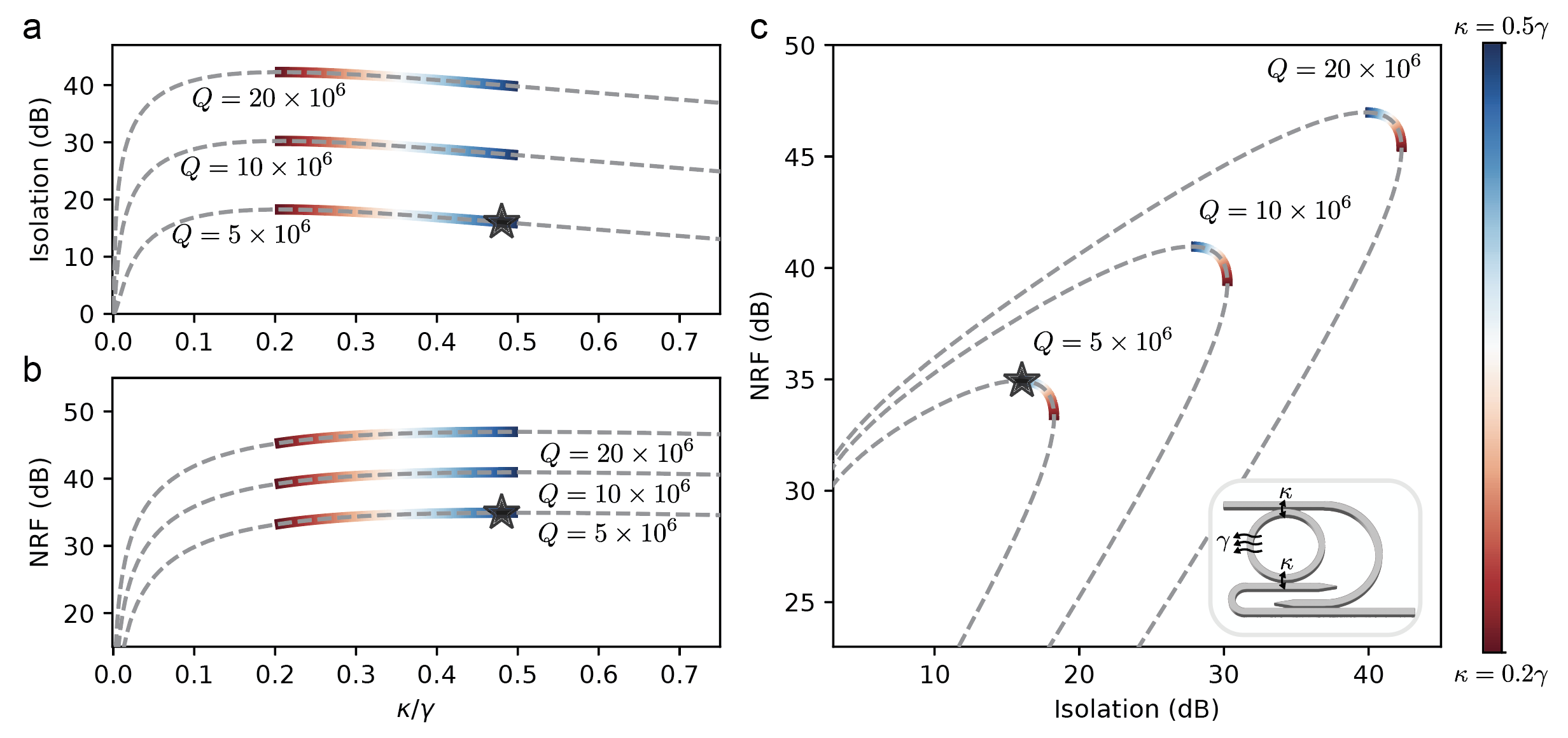}
\caption{{\bf Isolation/NRF Tradeoff.} 
{\bf a.} Theoretical isolation as a function of the ratio of ring coupling rate $\kappa$ to loss rate $\gamma$ for a range of state-of-the-art intrinsic Q factors where $n_2/n = \SI{1.2e-19}{m^2/W}$, $V_{\text{mode}} = \SI{1800}{\um^3}$.
{\bf b.} Theoretical noise reduction factor as a function of the ratio of ring coupling rate $\kappa$ to loss rate $\gamma$ where $Q_{\text{laser}} = \num{5e4}$, the laser amplitude phase coupling $\alpha = 2.5$, and return loss $\Gamma = \SI{20}{dB}$ when $\kappa = 0.5\gamma$.
{\bf c.} Isolation versus NRF tradeoff for a range of intrinsic Q factors.
Stars demarcate the values demonstrated in this work.
}
\label{fig:tradeoff}
\end{figure}

First, due to their different scaling, the isolation and noise reduction have different optimal loading of the ring. 
To maximize isolation, one should maximize $Q^2\eta = (\frac{1}{\kappa _1 + \kappa _2 + \gamma})^2 \frac{4\kappa_1(\kappa_2 + \gamma)}{(\kappa_1 + \kappa_2 + \gamma)^2}$, where $\kappa_1$ and $\kappa_2$ are the coupling rates of the ring to the two waveguides, and $\gamma$ is the intrinsic loss rate of the ring.
For $\kappa_1 = \kappa_2$, this leads to an optimum of $\kappa  = \frac{\sqrt{2}-1}{2}\gamma$, Fig 2a.
To maximize NRF, one should maximize $Q^2\Gamma = (\frac{1}{\kappa _1 + \kappa _2 + \gamma})^2 \frac{4 \kappa_1 \kappa_2}{(\kappa_1 + \kappa_2 + \gamma)^2}$, which with $\kappa_1 = \kappa_2$ leads to an optimum of $\kappa = 0.5\gamma$, Fig 2b, approximately 2.5 times more loaded than the isolation optimum. Outside of this range ($0.2\gamma < \kappa < 0.5\gamma)$ the isolation and noise reduction factors reduce together, Fig 2c.

Additionally, while mode volume does not directly affect the NRF, it does affect the thermorefractive noise (TRN) limit of the linewidth reduction (Supplemental Information Section 1). As the thermal limit of noise reduction scales linearly with the mode volume \cite{li2021reaching}, and the isolation scales inversely with the mode volume, there is a direct tradeoff between isolation and the TRN limit. As input power increases isolation without significantly affecting the NRF or TRN limit, the mode volume can thus be traded off with the input power. Finally, as the feedback strength is deterministic, it can be tuned with the coupling ratio of the directional coupler to trade off between maximum output power and noise reduction factor. Again, the noise reduction is ultimately limited by the TRN, and thus it can not be maximized arbitrarily. 

As every system has different noise, power, and isolation requirements, the optimal device configuration will likely be significantly different in every scenario. In this paper, we chose to load the devices to $\kappa = 0.5 \gamma$ (demarcated in Fig 2) and feed back 50\% of the out-coupled power to demonstrate a relatively low power device with good isolation and high NRF.

\section{High-Q Stability}

\begin{figure}[h!]
\centering\includegraphics[width=1\linewidth]{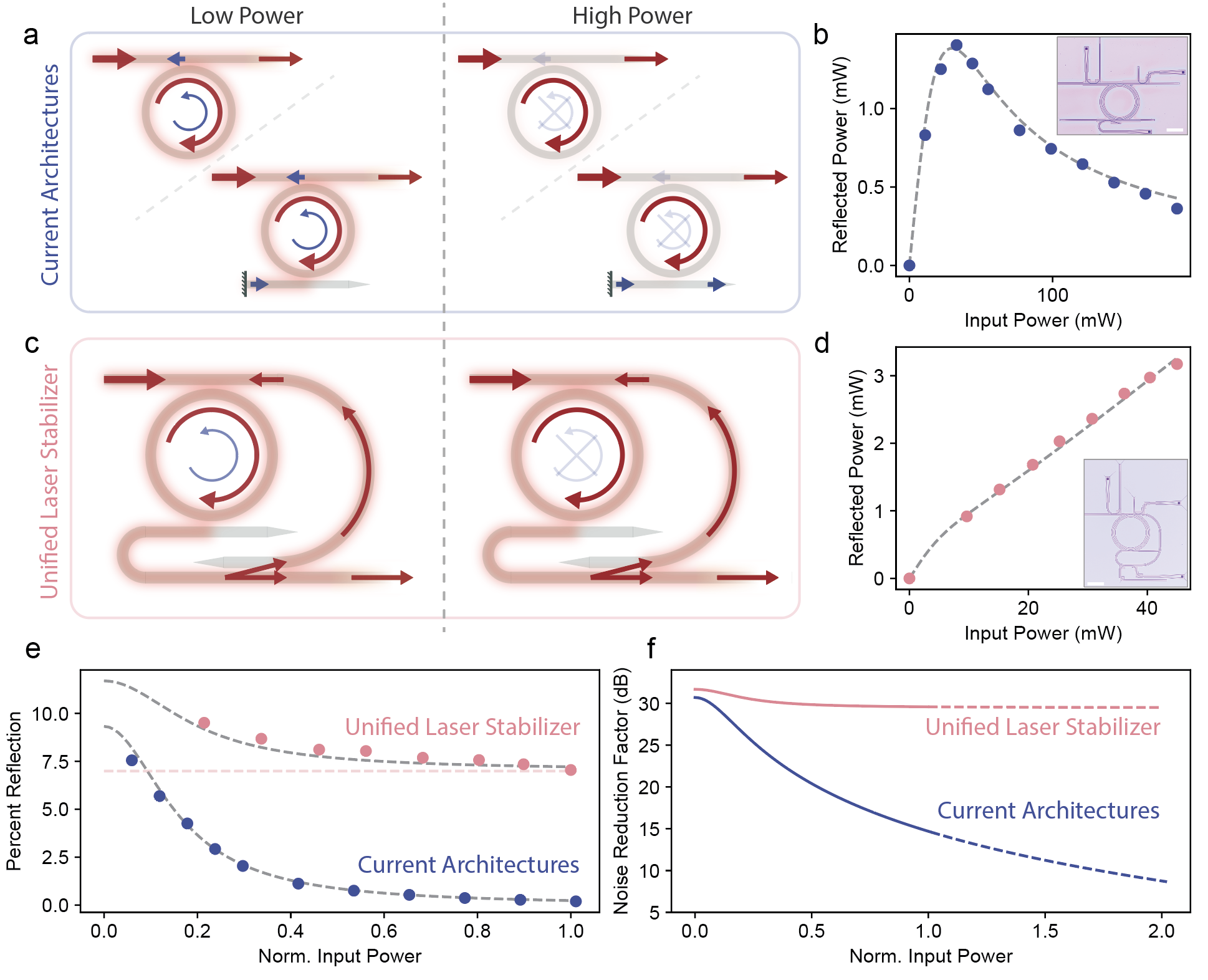}
\caption{{\bf High-Q Feedback.} 
{\bf a}. Current self-injection locking architectures based on parasitic or intentional back-scattering. Intrinsic ring back-scattering or a reflector are used to generate feedback signal. At low power, this works well, but as the power increases, the backscattered power no longer reaches the laser.
{\bf b}. Theoretical (dashed line) and experimental (data points) back-reflection as a function of input power corresponding to the architectures in (a). At low powers, the device is linear and exhibits a constant back-reflection ratio, but as power increases the nonlinear resonance splitting significantly attenuates the back-reflection. Inset shows device image with 100 $\mu$m scale bar. 
{\bf c}. Unified laser stabilizer feedback architecture. Significant feedback is provided back to the laser regardless of the input power.
{\bf d}. Theoretical (dashed line) and experimental (data points) back-reflection as a function of input power corresponding to the architecture in (c). The back-reflection strength is nearly independent of input power, with slight deviation arising from intrinsic back-scattering in the ring which is attenuated at higher powers. Inset shows device image with 100 $\mu$m scale bar. 
{\bf e}. Back-reflection as a percentage of input power for ULS in light red and current architectures in dark blue. Dashed pink line illustrates the asymptotic limit of the ULS reflection for high input powers. Due to different device quality factors, input power is normalized so that maximum input power corresponds to 15 dB isolation (45mW for ULS and 185mW for current architecture due to a lower Q factor). 
{\bf f}. Theoretical noise reduction factor vs. input power for each architecture assuming equivalent quality factors and insertion losses. Dashed lines show projection for higher power levels.}
\label{fig:Fig2}
\end{figure}

In addition to providing simultaneous isolation and noise reduction, the unified laser stabilizer topology solves a critical issue for self-injection locked lasers.
Current schemes for the self-injection locking of lasers rely on either parasitic back-reflection in rings \cite{xiang2021high, voloshin2021dynamics, jin2021hertz, guo2022chip, lihachev2022platicon, xiang20233d}, or a reflector placed after port 2 of the ring (drop port) \cite{shim2021tunable, corato2023widely}, Fig 3a. While these have been quite effective in the low power or low Q regimes, these schemes rely on the coupling and degeneracy of the clockwise and counterclockwise modes of the ring. If the power or Q was increased enough to break this degeneracy and provide isolation, the feedback strength would be heavily attenuated. As the noise reduction factor is proportional to the feedback strength, this attenuated power would lead to a dramatic reduction in efficacy of these systems (Supplemental Information Sections 2-3). Using published device parameters, we calculate at what power level and Q factor these state-of-the-art injection-locked lasers \cite{jin2021hertz, xiang2021high, lihachev2022platicon,  voloshin2021dynamics, lihachev2022low, corato2023widely, snigirev2023ultrafast, li2023high, ling2023self, clementi2023chip, shim2021tunable} will begin to fail (Supplemental Information Section 4). In many cases, these failure points are within an order of magnitude of the current operating point. 

In Fig 3b, we show the theoretical and experimental power-dependent back-reflection of a ring. Intrinsic back-scattering in the ring couples the clockwise and counterclockwise modes, causing a fraction of the input power to reflect. At low input power, the backreflected power increases linearly with the input power. As the input power increases, however, the clockwise and counterclockwise modes become increasingly detuned, and therefore less power is coupled backwards. This reduction is equal to the isolation ratio of the ring, here reaching a maximum of 15 dB. 

By instead providing nonreciprocal feedback through the ring, Fig 3c, we can achieve a nearly linear response even at high input powers. In Fig 3d, we show the backreflected power for a device with this new topology that also shows 15 dB of isolation. The reduced input power where this is achieved is due to a higher Q factor. At low input power, there are three competing effects -- the intrinsic ring back-scattering, the feedback through the ring, and the filtering the ring does to this feedback. At high power, however, the intrinsic back-scattering and feedback filtering are attenuated by the isolation ratio, allowing for a linear response (Supplemental Information Section 5). We compare the measured back-scattering ratio and the corresponding theoretical noise reduction factors of these topologies in Fig 3e-f. With current back-scattering based schemes, the presence of isolation directly degrades the noise reduction factor and locking range \cite{kondratiev2017self} (proportional to the isolation ratio and square root of the isolation ratio respectively), destroying its stabilization capability. In contrast, the unified laser stabilization topology preserves the high noise reduction factor across all power levels.

\section{Device Integration and Measurement}

We implement the ULS using a silicon nitride on insulator platform. While these devices are currently fabricated in-house, ultra high-Q silicon nitride photonics is fully compatible with foundry CMOS processes, as demonstrated by \cite{liu2021high, xiang20233d}, making scalable integration of these devices practical. To maintain a normal dispersion profile and prevent parasitic nonlinear conversion processes, we use a 310 nm waveguide height controlled by low-pressure chemical vapor deposition (LPCVD) of silicon nitride. To enable high quality factors (Supplemental Information Section 6), we use multi-mode rings ($\SI{4}{\um}$ width) and clad the devices with a thick oxide layer through plasma-enhanced chemical vapor deposition (PECVD).

\begin{figure}[h!]
\centering\includegraphics[width=1.02\linewidth]{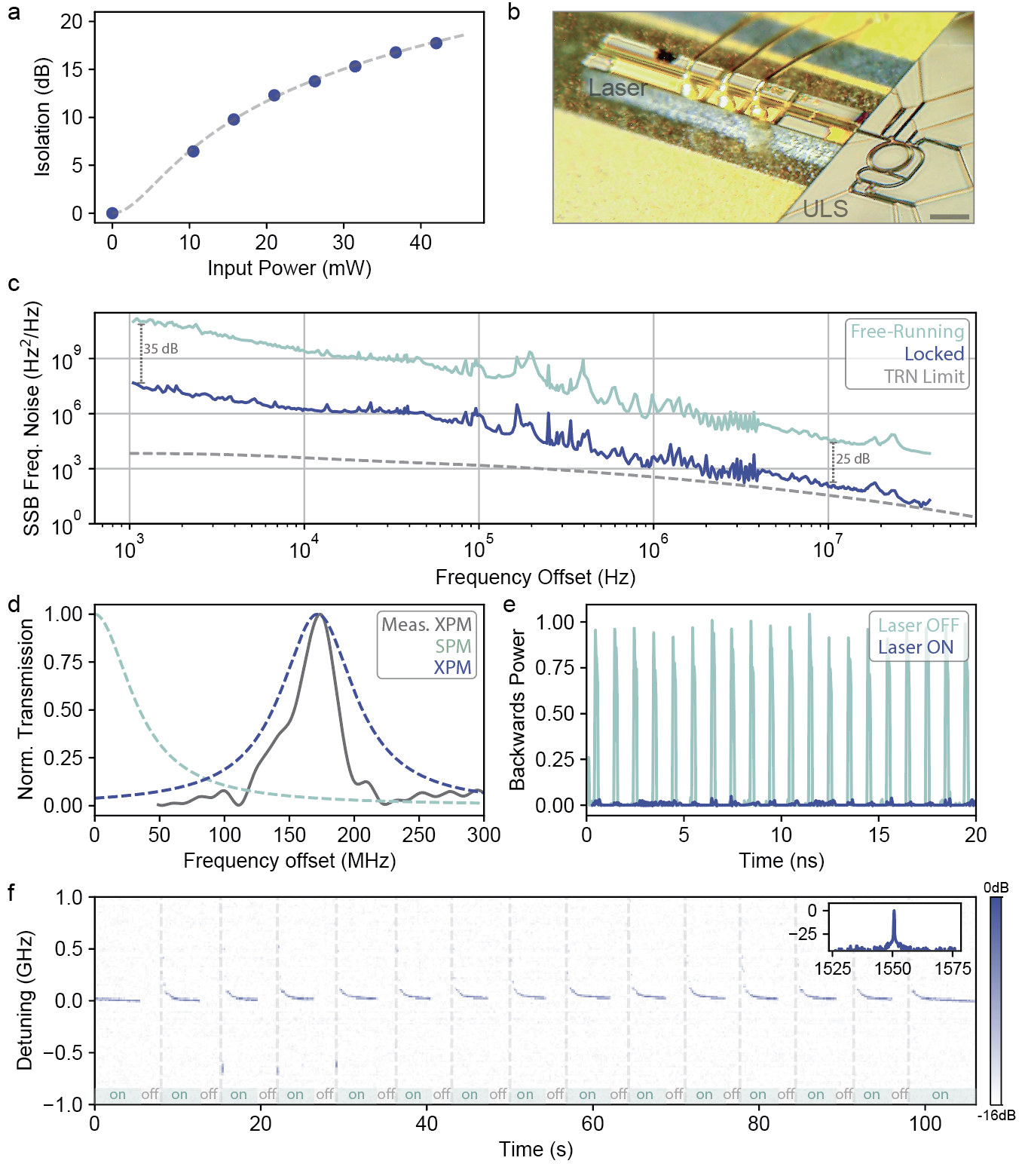}
\caption{{\bf Device Performance.} 
{\bf a}. Isolation vs. input power measured with a tunable external cavity laser.
{\bf b}. Device image showing hybrid integration of a DFB laser and ULS. Scale bar is $200~\mu m$.
{\bf c}. Single side-band frequency noise power spectral density of laser when free running (light green) and when locked to the ULS device (dark blue). Dashed line shows the simulated thermo-refractive noise of the device. 
{\bf d}. Measurement of the nonlinear splitting when locked to a semiconductor DFB laser with 33 mW of input power. Grey trace shows the normalized backward transmission of the device (measurement details in the methods section). Light green and dark blue dashed lines show the theoretical transmission spectrum of the self-phase and cross-phase} 
\label{fig:Fig1}
\end{figure}
\setcounter{figure}{3}

\begin{figure}[t!]
\caption{ cont. 
  resonances calculated from this measurement and from measurements of the ring linewidth. This splitting corresponds to an isolation ratio of 14 dB.   
{\bf e}. Measurement of backward transmission of 10 MHz pulses with the chip laser off (with the signal at the peak of the resonance) and with the chip laser on and locked to the ring (with the signal at the same frequency as the chip laser).
{\bf f}. Spectrograph of the heterodyne beat note between an isolated chip laser (operating in CW regime) and a tunable external cavity laser across numerous power cyclings. The laser current driver is turned on and off fully each cycle. Vertical lines correspond to the driver being switched on. Inset shows the corresponding optical spectrum.
}
\end{figure}

We first measure the isolation of the ULS with a tunable external cavity laser (ECL), Fig 4a. At each input power, we tune the laser across the ring resonance and record the backward transmission at the resonance peak. As expected, the isolation ratio follows a Lorentzian power-dependent curve proportional to the nonlinear detuning. We then edge-couple the ULS device to a semiconductor distributed feedback (DFB) laser that provides a coupled on-chip power of approximately 33 mW, Fig 4b. 
With a correlated self-heterodyne setup as described in \cite{yuan2022correlated}, we measure the single sideband (SSB) frequency noise power spectral density, Fig 4c. Compared to the free-running DFB laser, our hybrid device displays a noise reduction of 25-35 dB, limited at high offset frequency by the thermorefractive noise of the ring. At high frequency offset, close to the white noise floor of the laser, the SSB frequency noise power spectral density is reduced from 8200~Hz$^2$/Hz to 14~Hz$^2$/Hz, corresponding to instantaneous linewidths of 51~kHz and 89~Hz respectively (Supplemental Information Sections 7-9).


We then verify that isolation is achieved simultaneously with this reduction in frequency noise using two independent methods. First, we directly measure the nonlinear splitting of the ring modes, taking a similar approach to \cite{white2023integrated} modified to include a DFB pump.
To generate a probe signal, we take a tunable external cavity laser and reference its frequency with a heterodyne beat note to the output of the isolated DFB laser. We then scan the ECL across the resonance in the backward direction and monitor the transmission. To distinguish the probe transmission from the large feedback from the DFB, we modulate the probe and use a lock-in amplifier after photodetection (Supplemental Information Section 10). We then measure the frequency offset between the clockwise (resonant with the DFB) and counterclockwise (resonant with the ECL) modes using our heterodyne frequency reference. Fig 4d shows the trace from the lock-in amplifier, exhibiting a frequency splitting of 172~MHz. As the ring resonances have a half-width-half-maximum linewidth of 34.6~MHz (Supplemental Information Section 6), this splitting corresponds to an isolation of 14 dB, in agreement with the measurement in Fig 4a (assuming 33 mW input power). Second, we directly measure the transmission of pulses backwards through the ULS device with the DFB laser on and off, Fig 4e. To do this, we ensure the ECL probe is at the same frequency as the DFB pump using their heterodyne beat note. As the probe and pump are at the same frequency, we extract the transmission from the DC (DFB feedback) and AC (ECL pulse and DFB feedback heterodyne) components. We again find an isolation ratio in agreement with Fig 4a and 4d.


In this system, the combination of thermal locking and self-injection locking allows for laser stability without external feedback loops. In Fig 4f, we demonstrate the robustness of this type of laser system: when the laser is locked, it can be fully turned on and off over the timescale of at least tens of seconds. Without the self-injection locking feedback, the ring would need to be thermally tuned each time upon startup in order to align its transmission spectrum with the laser.
Finally, we ensure that this topology is still capable of taking advantage of the rich dynamics that traditionally self-injection locked lasers exhibit \cite{shen2020integrated, jin2021hertz, li2023high, lihachev2022platicon}. The control of feedback phase is a critical feature to enable or to avoid the comb formation when self-injection locked \cite{lihachev2022platicon} (Supplemental Information Section 11). By tuning the coupling gap and thus feedback phase with a piezoelectric positioner, we can observe both CW and frequency comb operation, and even second harmonic generation (Supplemental Information Sections 11-12) \cite{clementi2023chip}.

\section{Conclusion}

Here we have demonstrated the first fully on-chip narrow-linewidth laser system, complete with built-in optical isolation. To do this, we combined self-injection locking and isolation in a single CMOS-compatible silicon nitride unified laser stabilizer. While previous self-injection schemes function well at low power, we show that they fail to function at higher powers. We propose a device topology that provides strong feedback independent of input power while simultaneously providing isolation. Our fabricated devices passively isolate an integrated DFB laser by 14 dB and at the same time reduce its frequency noise by 25-35 dB, and operate with turn-key reliability. By increasing the device quality factor through commercial scale fabrication \cite{liu2021high}, the isolation and noise reduction can be even further enhanced, and the power threshold for operation can be reduced (Fig 2 and Supplemental Information Section 13). While we demonstrated a ULS that stabilizes a laser through injection locking, a ULS could instead be used as one of the mirrors inside a laser cavity itself. This would lead to a lower insertion loss, enabling even higher isolation and narrower linewidths.
As many hybrid and heterogeneously integrated photonics already include low-loss silicon nitride or similar materials, these devices can be readily integrated into state-of-the-art systems for sensing, metrology, and quantum technologies.

\clearpage

\section*{Methods}

\subsection*{Device Fabrication}

Silicon nitride thin films (310nm) were deposited on silicon dioxide/silicon wafers with low-pressure chemical vapor deposition. Devices were then patterned with ZEP520A resist and e-beam lithography (JEOL JBX-6300FS). After development, the patterns were transferred to the silicon nitride using inductively coupled plasma etching with CHF3/CF4 chemistry. Devices were then cleaned with a Piranha solution, annealed at 1,100$^{\circ}$ C in a nitrogen environment, and oxide clad using spun hydrogen silsesquioxane (HSQ) and thermal oxide. The devices were annealed a final time under the same conditions and laser stealth diced to create clean facets.

\subsection*{Frequency splitting measurement}

To measure the frequency splitting of the clockwise and counterclockwise ring modes when pumped with the DFB laser, we implement the setup shown in Supplemental Fig 11. We couple the DFB laser (PhotonX) directly to the chip and use a grating coupler and fiber-coupled circulator to extract the laser power from the output. We then beat this output with a tap-off from a tunable ECL (Toptica) on a photodiode and use the beat note to tune the ECL to frequency degeneracy with the DFB. We can then scan the tunable laser frequency over time and track the frequency-dependent transmission. To recover the frequency-dependent transmission of the ECL over the large amount of backward-going DFB power, we modulate the ECL with an electro-optic modulator (EOM) and use a lock-in amplifier to read out the transmission. We note that the transmission at low offset frequency is not shown in the main figure set (Fig 4d) for clarity as it contains heterodyne noise that does not represent transmission.

\subsection*{Pulsed isolation measurement}

To measure the backward transmission of a pulsed backward signal, we implement the setup shown in Supplemental Fig 12. Here we also use the heterodyne beat note between the DFB laser and the tunable ECL to tune the ECL to the same frequency as the DFB. We then modulate the ECL with 10MHz pulses using an EOM and monitor the backward transmission without tuning the ECL. As the backward transmission is at the same frequency as the DFB pump, the photodiode generates a beat note at the modulation frequencies. We measure this beat note by amplifying the photodiode output with a transimpedance amplifier and high pass filtering the resulting signal to remove low frequency and flicker noise. We then record both the beat note and the much larger DC component (with high pass filter removed). Using the beat note and DC component, we can back-calculate the original signal. To compare the transmission to the un-pumped ('laser off') transmission, we do the same measurement without the DFB laser, ensuring that the ECL is tuned onto resonance with the ring resonator.
We note that the noise floor of the 'laser off' measurement in Fig 4e is clipped off to not obscure 'laser on' data.

\subsection*{Thermorefractive Noise Simulations}

To verify whether the frequency noise of our device reached the thermal limit, we closely followed the work by \cite{kondratiev2018thermorefractive, huang2019thermorefractive} to simulate thermorefractive noise (TRN) and corresponding resonance frequency fluctuations in the ring resonator. Thermal noise was found using fluctuation-dissipation theorem (FDT) and finite-element method (COMSOL Multiphysics). Specifically, we added a harmonic perturbative heat source in the system with the same spatial distribution as the fundamental waveguide eigenmode, solved the heat transfer equation in the frequency domain, calculated the dissipated heat energy during one cycle, and used FDT to convert dissipated heat energy to frequency fluctuations. In the simulations, we assumed $4000\times \SI{310}{nm}$ Si$_3$N$_4$ waveguide ($R=\SI{100}{\um}$) surrounded by SiO$_2$ cladding ($\SI{2.5}{\um}$ on top, $\SI{5}{\um}$ on bottom) with air above and Si substrate below. The material properties used in the simulation and a more detailed analysis of TRN (temperature dependency, effects of device geometry) are described in Supplemental Information Section 1.
\\

\noindent\textbf{Data Availability}\\
\noindent 
All data are available from the corresponding authors upon reasonable request.
\\

\noindent\textbf{Additional Information}\\
\noindent 
Supplementary Information is available for this paper. Correspondence and requests for materials should be addressed to adwhite@stanford.edu, gahn@stanford.edu.
\\

\noindent\textbf{Acknowledgments}\\
\noindent 
 A.W. acknowledges the Herb and Jane Dwight Stanford Graduate Fellowship (SGF) and the NTT Research Fellowship. G.H.A. acknowledges support from STMicroelectronics Stanford Graduate Fellowship (SGF) and Kwanjeong Educational Foundation. Authors from Stanford University and UCSB acknowledge funding support from DARPA under the LUMOS program. Authors from University of Washington acknowledge funding support from NSF under NSF-QII-TAQS-1936100. Part of this work was performed at the Stanford Nano Shared Facilities (SNSF)/Stanford Nanofabrication Facility (SNF), supported by the National Science Foundation under award ECCS-2026822.
\\

\noindent\textbf{Author Contributions}\\
\noindent 
A.D.W., G.H.A., and K.V.G. conceived of the project. A.D.W., G.H.A., R.L., and K.V.G. performed the experiments. G.H.A. developed the silicon-nitride fabrication process and fabricated the devices with assistance from A.S. and A.M.. J.G., T.J.M., L.C. and J.E.B. provided the semiconductor laser chip and experimental guidance. J.V. supervised the project. All authors contributed to data analysis and writing of the manuscript.
\\

\noindent\textbf{Competing Interests}\\
\noindent 
A.D.W., G.H.A., K.V.G., and J.V. have filed a patent application for the ULS laser architecture (PCT/US2023/032287).


\bibliography{References}

\end{document}